\documentstyle[aps,epsf,rotate,multicol]{revtex}

\begin{document}

\draft

\title{Scaling for the Percolation Backbone}

\author{M.~Barth\'el\'emy{\footnote{Permanent address: CEA-BIII,
Service de Physique de la Mati\`ere Condens\'ee, France.}}$^{1}$,
S.~V.~Buldyrev$^{1}$, S.~Havlin$^{2}$ and H.~E.~Stanley$^{1}$}

\address{ $^1$ Center for Polymer Studies and Dept. of Physics,
		Boston University, Boston, MA 02215 \\
	 $^2$ Department of Physics, Bar-Ilan University, Ramat-Gan
        52900, Israel}

%\date{\today}

\maketitle
\begin{abstract}
We study the backbone connecting two given sites of a two-dimensional
lattice separated by an arbitrary distance $r$ in a system of size $L$.
We find a scaling form for the average backbone mass: $\langle
M_B\rangle\sim L^{d_B}G(r/L)$, where $G$ can be well approximated by a
power law for $0\le x\le 1$: $G(x)\sim x^{\psi}$ with $\psi=0.37\pm
0.02$. This result implies that $\langle M_B\rangle \sim
L^{d_B-\psi}r^{\psi}$ for the entire range $0<r<L$. We also propose a
scaling form for the probability distribution $P(M_B)$ of backbone mass
for a given $r$. For $r\approx L$, $P(M_B)$ is peaked around $L^{d_B}$,
whereas for $r\ll L$, $P(M_B)$ decreases as a power law,
$M_B^{-\tau_B}$, with $\tau_B\simeq 1.20\pm 0.03$. The exponents $\psi$
and $\tau_B$ satisfy the relation $\psi=d_B(\tau_B-1)$, and $\psi$ is
the codimension of the backbone, $\psi=d-d_B$.
\end{abstract}

\vskip 1cm

%64,60Ak=RG, fractal percolation
%0545df=fractals in General

\begin{multicols}{2}

%%%%%%%%%%%%%%%%%%%%%%INTRODUCTION

The percolation problem is a classical model of phase transitions, as
well as a useful model for describing connectivity phenomena, and in
particular for describing porous media
\cite{Bunde96,Stauffer92,Sahimi92}. At the percolation threshold $p_c$,
the mass of the largest cluster scales with the system size $L$ as
$M\sim L^{d_f}$. The fractal dimension $d_f$ is related to the space
dimension $d$ and to the order parameter and correlation length
exponents $\beta$ and $\nu$ by $d_f=d-\beta/\nu$
\cite{Bunde96,Stauffer92,Sahimi92}. In two dimensions, $d_f=91/48$ is
known exactly.

An interesting subset of the percolation cluster is the backbone which
is obtained by removing the non-current carrying bonds from the
percolation cluster\cite{Kirkpatrick78}. The structure of the backbone
consists of blobs and
links\cite{Bunde96,Herrmann84,Skal74,deGennes76}. The backbone can in
fact be further partitioned into subsets according to the magnitude of
the electric current carried\cite{Rammal85,ARC87}. The backbone is
relevant to transport properties \cite{Bunde96,Stauffer92,Sahimi92} and
fracture \cite{Roux90}. The fractal dimension $d_B$ of the backbone can
be defined via its typical mass $M_B$, which scales with the system size
$L$ as $M_B\sim L^{d_B}$. The backbone dimension is an independent
exponent and its exact value is not known. A current numerical estimate
\cite{Grass99} is $d_B=1.6432\pm 0.0008$.

The operational definition of the backbone has an interesting
history\cite{Bunde96,Stauffer92,Sahimi92}. Customarily, one defines the
backbone using parallel bars, and looks for the percolation cluster (and
the backbone) which connects the two sides of the
system\cite{Kirkpatrick78}. A different situation arises in oil field
applications \cite{Lee98}, where one studies the backbone connecting two
wells separated by an {\it arbitrary} distance $r$. This situation is
important for transport properties, since in oil recovery one injects
water at one point and recovers oil at another point \cite{Lee98}. From
a fundamental point of view, it is important to understand how the
percolation properties depend on different boundary conditions.

We study here the backbone connecting two points separated by an
arbitrary distance $r$ in a two-dimensional system of linear size $L$.
One goal is to understand the distribution of the backbone mass
$M_B(r,L)$, and how its average value scales with $r$ and $L$ in the
entire range $0< r < L$.

%%%%%%%%%%%%%%%%%%%%%%%%%%%%%%%%

We choose two sites $A$ and $B$ belonging to the infinite percolating
cluster on a two-dimensional square lattice (the fraction of bonds is
$p=p_c=1/2$). $A$ and $B$ are separated by a distance $r$ and
symetrically located between the boundaries\cite{Herrmann84.2}. Using
the burning algorithm, we determine the backbone connecting these two
points for values of $L$ ranging from $100$ to $1000$. For each value of
$L$, we consider a sequence of values of $r$ with $2\le r\le L-2$. In
order to test the universality of the exponents, we perform our study on
three lattices: square, honeycomb and triangular lattice. For
simplicity, we restrict our discussion here to the square lattice, as we
find similar results for the other two lattices.

We begin by studying the backbone mass probability distribution
$P(M_B)$. We show that $P(M_B)$ obeys a simple scaling form in the
entire range of $r/L$,
\begin{equation}
\label{pscal}
P(M_B)\sim \frac{1}{r^{d_B}}F\left(\frac{M_B}{r^{d_B}}\right),
\end{equation}
where $F(x)$ is a scaling function, whose shape depends on the ratio $r/L$. 

For $r\approx L$, it seems reasonable to assume that $P(M_B)$ will be
peaked around its average value $<M_B>\sim L^{d_B}$. The data collapse
predicted by Eq.~(\ref{pscal}) is represented in Fig.~1(a). In this case, the
scaling function $F$ is peaked at approximately $L^{d_B}$.

However, the case $r\ll L$ is far less clear.  In fact, we expect for
$r\ll L$ that the backbone mass fluctuates greatly from one realization
to another, since its minimum value can be $r$ and its maximum can be of
order $L^{d_f}$. Fig.~1(b) shows a log-log plot of $P(M_B)$.  It has a
lower cut-off of order $r$ (since the backbone must connect points $A$
and $B$) and a upper cut-off of order $L^{d_B}$. We find good data
collapse (Fig.~1(c)), which indicates that the scaling function $F$ is a
power law in the range from $r^{d_B}$ to $L^{d_B}$, with exponent
approximately $\tau_B\simeq 1.20\pm 0.03$ (there is a cut-off at
$M_B\sim L^{d_B}$ not shown here). The exponent $\tau_B$ is connected to
the blob size distribution \cite{Herrmann84} since typically, the two
sites belong to the same blob, and the sampling of backbones is
equivalent to sampling of the blobs. From \cite{Herrmann84},
\begin{equation}
\label{taub}
\frac{d}{d_B}=\tau_B.
\end{equation}
This relation gives the estimate $\tau_B\simeq 1.22$ in good agreement
with our numerical simulation.

We note that for larger values of $M_B$, a ``bump'' (indicated by an arrow
on Fig.~1(b)) located at approximately $L^{d_B}$ appears and assumes
increasing importance when $r$ approaches $L$.

We now study the average backbone mass $\langle M_B\rangle$. From
dimensional considerations, the $r$ dependence can only be a function of
$r/L$. We thus propose the following {\it Ansatz}:
\begin{equation}
\label{mscal}
\langle M_B(r,L)\rangle=L^{d_B}G\left(\frac{r}{L}\right).
\end{equation}
In Fig.~2(a), we show a double logarithmic scale $M_B$ versus $r$ for
different values of $L$. In order to test the Eq.~(\ref{mscal}), we scale
the data of Fig.~2(a). The data collapse is obtained using $d_B=1.65$
and is shown on Fig.~2(b). This (log-log) plot supports the scaling
Ansatz (\ref{mscal}). Moreover, one can see that the scaling function
$G$ is, surprisingly, a pure power law on the entire range $[0,1]$, with
exponent $\psi=0.37\pm0.02$.

%%%%%%%%%%%%%%%%%%%%%%%%%%%%%%

The results (\ref{pscal}) and (\ref{mscal}) are consistent, since if
(\ref{pscal}) holds with a power law behavior for the scaling function
$F(x)\sim x^{-\tau_B}$ for $x>1$, and $F(x)=0$ for $x<1$, then the
average mass is given by
\begin{equation}
\label{mscal2}
\langle M_B(r,L)\rangle=\int^{L^{d_B}}_{r}F\left(\frac{M}{r^{d_B}}\right) 
\frac{dM}{r^{d_B}}M. 
\end{equation}
Assuming that $L/r$ is large enough, the integral in (\ref{mscal2}) can
be approximated as $L^{d_B-\psi}r^\psi$, where
\begin{equation}
\label{psi}
\psi=d_B(\tau_B-1)
\end{equation}
In our simulation $\tau_B\approx 1.20\pm 0.03$, which leads to the value
$\psi\approx 0.33\pm 0.05$ in reasonable agreement with the value
measured directly on the average mass.

Moreover, using Eq.~(\ref{taub}) together with Eq.~(\ref{psi}), we obtain
\begin{equation}
\psi=d-d_B
\end{equation}
which means that $\psi$ is the codimension of the fractal backbone.

To summarize, we find that for any value of $r/L$, the scaling form,
Eq.~(\ref{pscal}), for the probability distribution is valid. The shape
of the scaling function $F$ depends on $r/L$, being a peaked
distribution for $r\approx L$, and a power law for $r\ll L$. The average
backbone mass varies with $r$ and $L$ according to Eq.~(\ref{mscal2}).
For fixed system size, it varies as $\langle M_B\rangle \simeq r^{\psi}$
(for $0<r<L$). The value of $\psi$ is small ($\psi\approx 0.37$)
indicating that the backbone mass does not change drastically as $r$
changes. On the other hand, the exponent governing the variation of
$\langle M_B\rangle $ with $L$ for fixed $r$ is expected to be larger, 
with $\langle M_B\rangle \sim r^{d_B-\psi}$. This exponent $d_B-\psi$ is
not equal to the fractal dimension $d_B$ of the backbone, but is smaller
by an amount equal to $\psi$.

\vskip 1cm

Acknowledgements. We would like to thank P. Gopikrishnan for help with
the simulations and L. A. N.~Amaral, A.~Coniglio, N. V.~Dokholyan,
Y.~Lee, P. R.~King and V.~Plerou for stimulating discussions. MB thanks
the DGA for financial support. The Center for Polymer Studies is
supported by the NSF and BP Amoco.

%%%%%%%%%%%%%%%%%%%%%%%%%%%%%%%%%%%%%%%%%%%%%%%%%%%%%%%%%%%% References

%%%%%%%%%%%%%%%%%%%%%%%%%%%%%%%%%%%%%%%%%%%%%%%%%%%%%%%%%%% Figures

%\end{multicols}

\newpage

\begin{figure}
\narrowtext
\centerline{
\epsfysize=0.8\columnwidth{\rotate[r]{\epsfbox{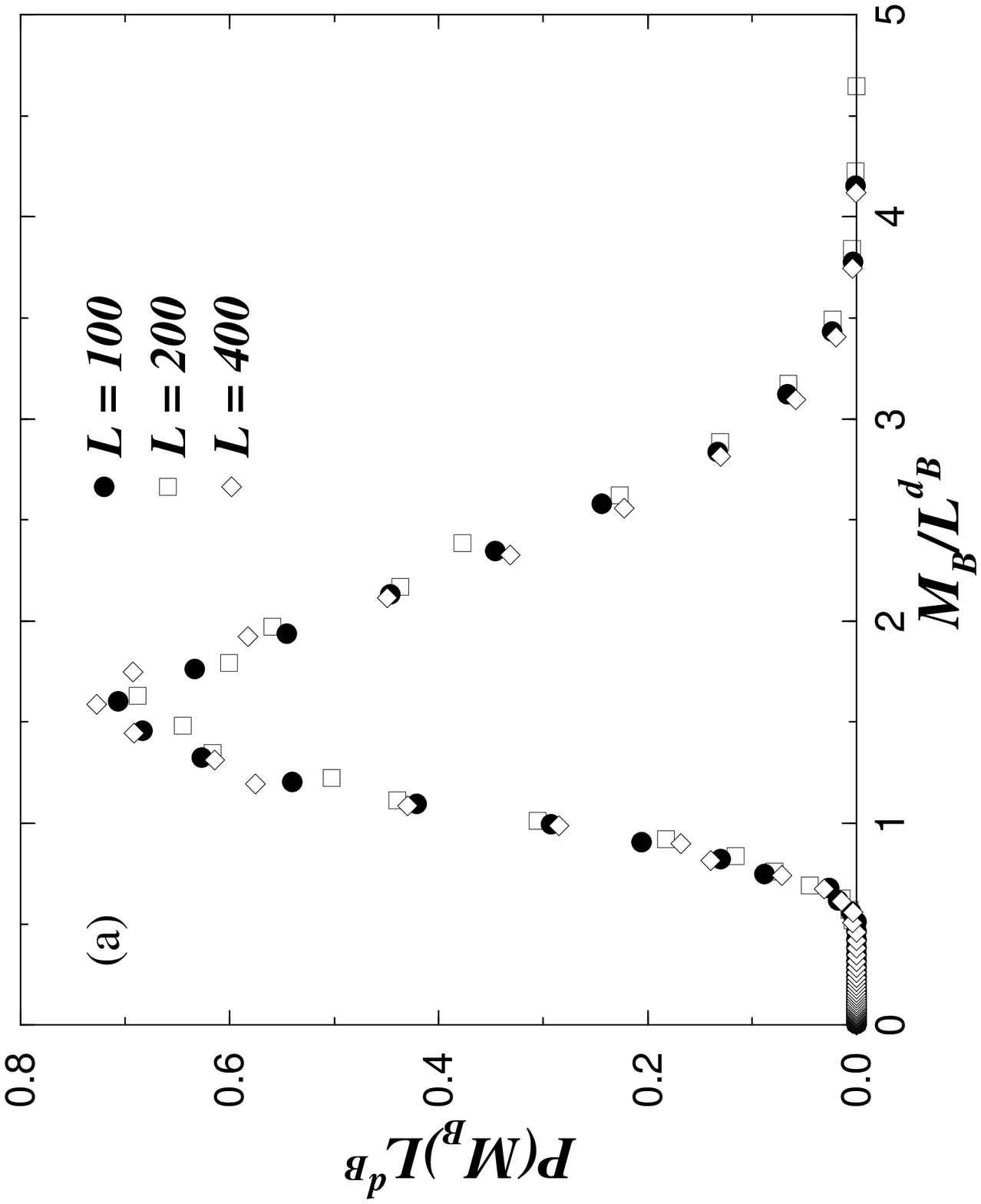}}}
}
\centerline{
\epsfysize=0.8\columnwidth{\rotate[r]{\epsfbox{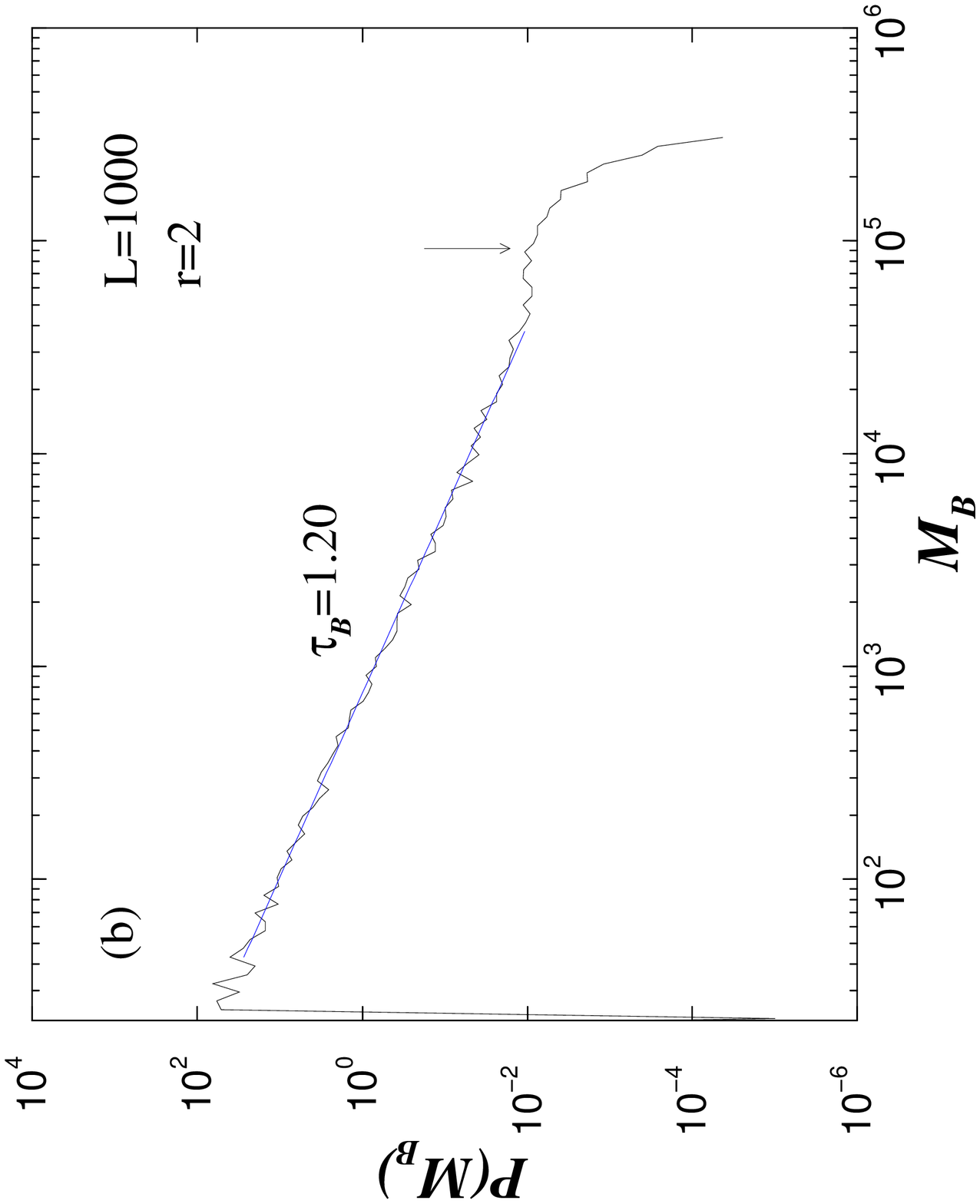}}}
}
\centerline{
\epsfysize=0.8\columnwidth{\rotate[r]{\epsfbox{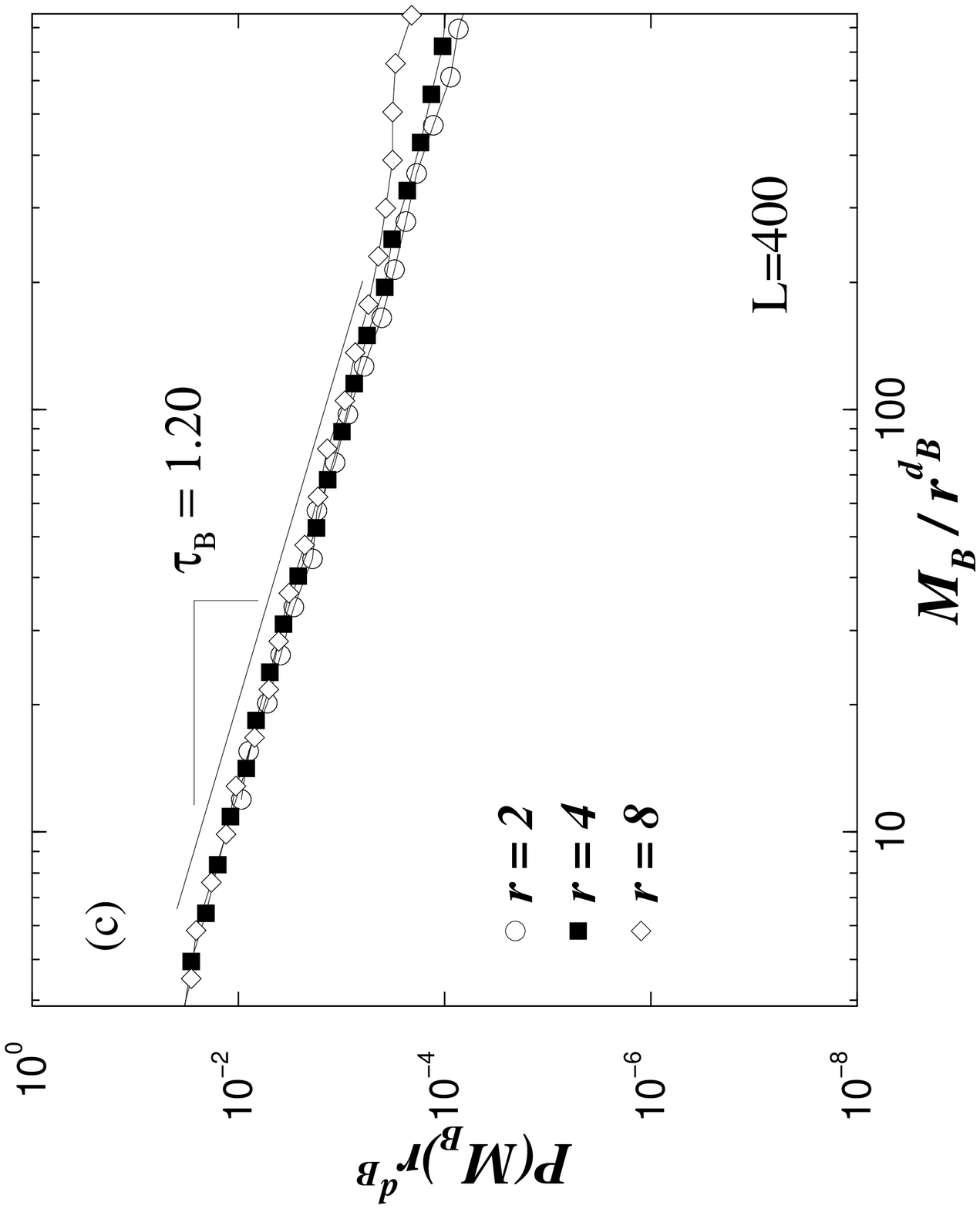}}}
}
\vspace*{0.1cm}
\caption{ (a) Data collapse of $P(M_B)$ using Eq.~(1) for three
different value of $r\simeq L$. (b) Probability distribution of the
backbone mass for $L=1000$ and $r=2$ (computed with $10^5$
configurations). The exponent $\tau_B$ is obtained by a linear fit over
the range $30<M_B<3\times 10^4$ and the error bar on $\tau_B$ is around
$0.03$. The arrow denotes the fact that $M_B$ peaks as $L^{d_B}$. (c)
Data collapse of $P(M_B)$ for $L=400$ using Eq.~(1) for three different
values of $r$. }
%\label{Fig.~pM}
\end{figure}

%%%%%%%%%%%%%%%%%%%%%%%%%%%

\begin{figure}
\narrowtext
\centerline{
\epsfysize=0.8\columnwidth{\rotate[r]{\epsfbox{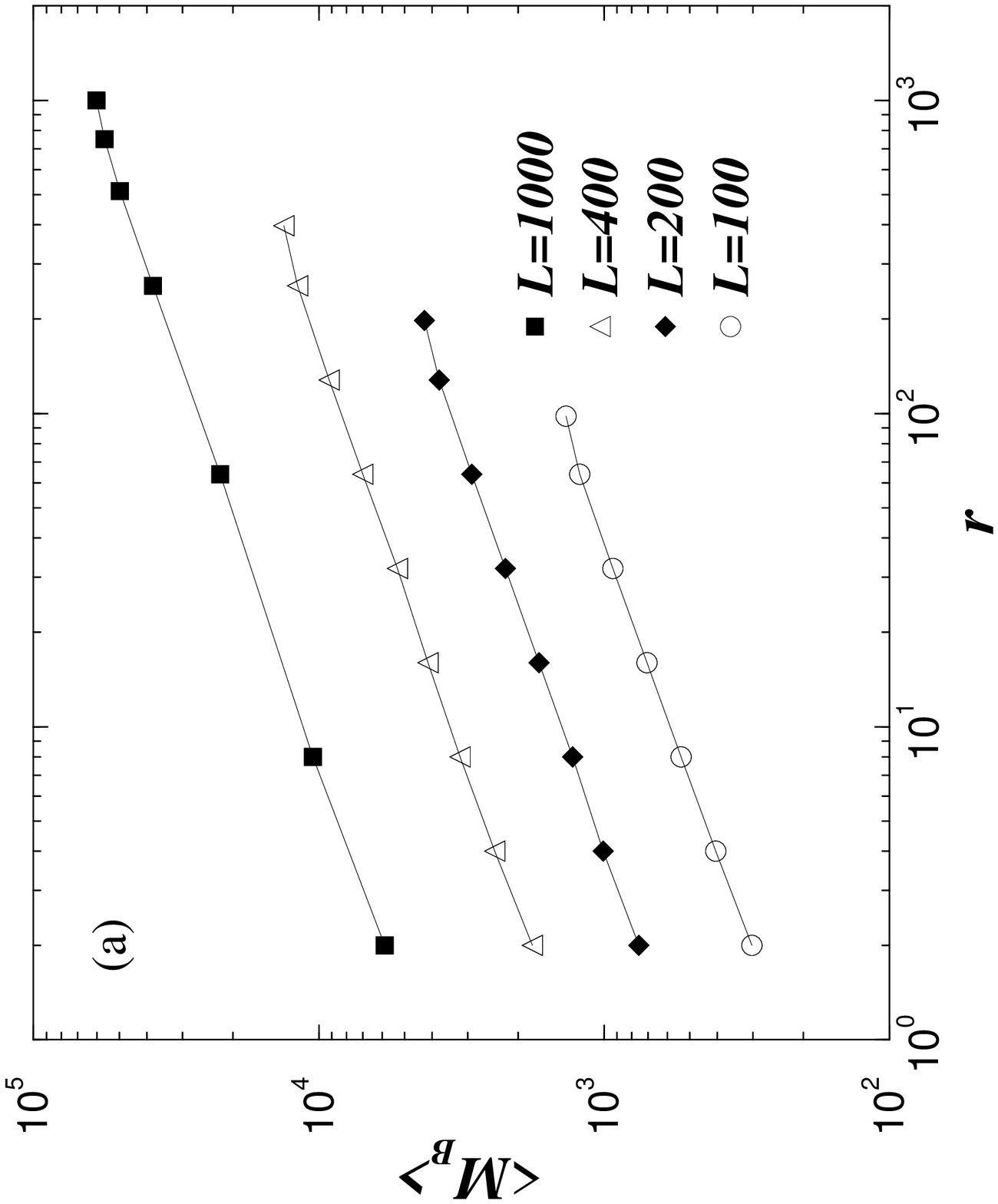}}}
}
\centerline{
\epsfysize=0.8\columnwidth{\rotate[r]{\epsfbox{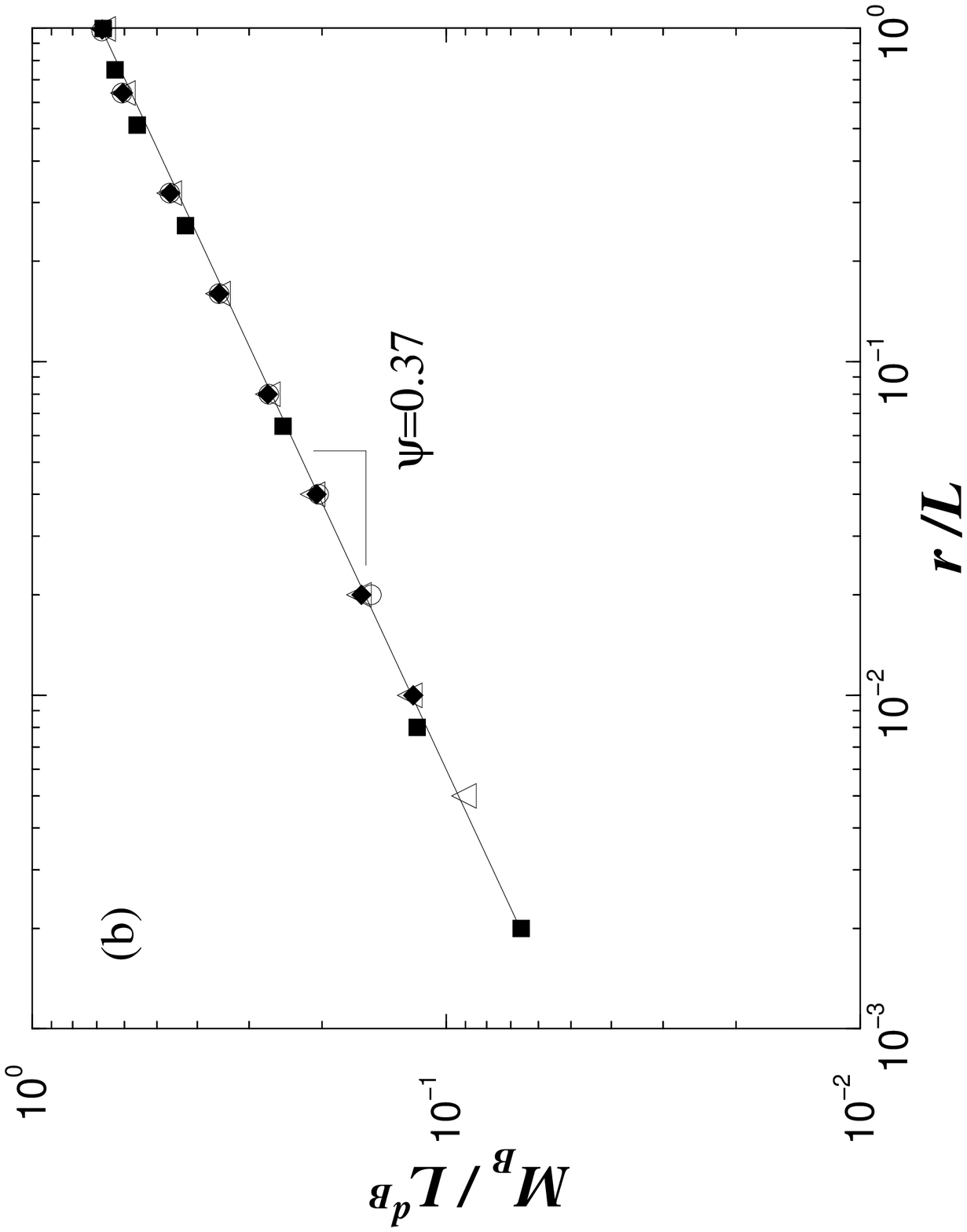}}}
}
\vspace*{0.1cm}
\caption{ 
(a) Log-log plot of the average backbone mass $\langle M_B\rangle $
versus $r$ for four different values of $L$. (b) Data from Fig.~2(a)
collapsed with the use of the scaling form proposed in Eq.~(2). The
error on $\psi$ is typically $0.02$.}
%\label{fig1}
\end{figure}

%\vfill
%\eject
%\begin{multicols}{2}

%%%%%%%%%%%%%%%%%%%%%%%%%%%

\end{multicols}


\begin{references}

\bibitem{Bunde96}
A.~Bunde and S.~Havlin (eds.), {\it Fractals and Disordered Systems,
Second Edition\/} (Springer, Berlin, 1996), and refs. therein.

\bibitem{Stauffer92}
D. Stauffer and A.~Aharony, {\it Introduction to Percolation Theory\/}
(Taylor and Francis, London, 1992).

\bibitem{Sahimi92}
M.~Sahimi, {\it Applications of Percolation Theory\/} (Taylor and
Francis, London, 1992).

\bibitem{Kirkpatrick78} 
S.~Kirkpatrick, {\it AIP Conf. Proc.} {\bf 40}, 99 (1978); G. Shlifer,
W. Klein, P. J. Reynolds, and H. E. Stanley, {\it J.  Phys. A\/} {\bf 12},
L169--174 (1979).

\bibitem{Herrmann84} 
H. J.~Herrmann and H. E.~Stanley, {\it Phys. Rev. Lett.} {\bf 53}, 1121
(1984).

\bibitem{Skal74} 
H. E. Stanley, {\it J.  Phys. A\/} {\bf 10}, L211--220 (1977).

\bibitem{deGennes76} 
A. Coniglio, {\it Phys. Rev. Lett.} {\bf 46}, 250--253 (1981).

\bibitem{Rammal85} R.~Rammal, C.~Tannous, P.~Breton and A.M.S. Tremblay,
{\it Phys. Rev. Lett} {\bf 54}, 1718 (1985).

\bibitem{ARC87} 
L.~de Arcangelis, A.~Coniglio and S.~Redner, {\it Phys. Rev. B\/} {\bf
36}, 5631 (1987).

\bibitem{Roux90} 
S.~Roux and H. J.~Herrmann (eds.), {\it Statistical Models for the
Fracture of Disordered Media\/} (North-Holland, Amsterdam, 1990).

\bibitem{Grass99} 
P.~Grassberger, {\it Physica A} {\bf 262}, 251 (1999).

\bibitem{Lee98} 
Y.~Lee, J. S.~Andrade Jr, S. V. Buldyrev, S. Havlin, P. R. King, G.
Paul, and H. E. Stanley, preprint cond-mat 9903066.

\bibitem{Herrmann84.2} 
H. J.~Herrmann, D.~Hong and H. E.~Stanley, {\it J. Phys. A\/} {\bf 17},
L261 (1984).

\end{references}
\end{document}